\documentclass[pra,twocolumn,showpacs,showkeys,amsmath,nobalancelastpage,floatfix,preprintnumbers,amssymb]{revtex4}

\usepackage[T1]{fontenc}
\usepackage[latin1]{inputenc}
\usepackage[english]{babel}

\usepackage{graphicx}
\usepackage{bm}

\newcommand{\ket}[1]{\ensuremath{\left| #1 \right \rangle}}
\newcommand{\bra}[1]{\ensuremath{\left \langle #1 \right |}}
\newcommand{\UCU}[2]{F^{#1}_{#2}\left(U\!\left(2\right)\right)}
\newcommand{\UC}[3]{F^{#1}_{#2}\left( #3 \right)}
\newcommand{\TUCU}[2]{\tilde{F}^{#1}_{#2}\left(U\!\left(2\right)\right)}
\newcommand{\CNOT}[2]{C^{#1}_{#2}}

\begin{document}

\title{Quantum circuits with uniformly controlled one-qubit gates}
\author{Ville Bergholm}
\email{vberghol@focus.hut.fi}
\author{Juha J. Vartiainen}
\author{Mikko M\"ott\"onen}
\author{Martti\ M.\ Salomaa}
\affiliation{Materials Physics Laboratory, POB 2200 (Technical Physics)\\
FIN-02015 HUT, Helsinki University of Technology, Finland}
\date{\today}

\begin{abstract}
Uniformly controlled one-qubit gates are quantum gates which can be
represented as direct sums of two-dimensional unitary operators
acting on a single qubit.
We present a quantum gate array which implements any $n$-qubit gate of
this type using at most $2^{n-1} - 1$ controlled-NOT gates, $2^{n-1}$ one-qubit gates
and a single diagonal $n$-qubit gate.
The circuit is based on the so-called quantum multiplexor, for which we
provide a modified construction.
We illustrate the versatility of these gates by applying them to the
decomposition of a general $n$-qubit gate and a local
state preparation procedure. Moreover, we study their implementation
using only nearest-neighbor gates.
We give upper bounds for the one-qubit and controlled-NOT gate counts for all the
aforementioned applications.
In all four cases, the proposed circuit topologies either improve on or achieve the
previously reported upper bounds for the gate counts.
Thus, they provide the most efficient method for general gate decompositions
currently known.
\end{abstract}


\maketitle

\parindent 0mm
\parskip 5mm

\section{Introduction}

A quantum computer is an emerging computational device based on
encoding classical information into a quantum-mechanical
system~\cite{NielsenChuang}. Since the breakthrough factorization
algorithm by Shor in 1994~\cite{shor94},
progress in research on quantum computing has been expeditious~\cite{galindo}.
Most quantum computers involve a collection of two-level
systems, a quantum register, in which the information is stored.
The two-level systems themselves, called qubits, can also be
replaced by arbitrary $d$-level systems, known as qudits~\cite{bullock_quditl}.
The computation is performed by the unitary temporal evolution of the
register, followed by a measurement.
In order to execute the desired algorithm, one has to be able to exert
sufficient control on the Hamiltonian of the register to obtain the
required propagators.
These unitary propagators, acting on the register, are
called quantum gates.

The current paradigm for implementing quantum algorithms is the
quantum circuit model~\cite{deutsch}, in which the algorithms
are compiled into a sequence of simple gates acting on one or
more qubits. The detailed decomposition of an arbitrary quantum gate into
an array of elementary gates was first presented by Barenco et
al.~\cite{barenco}. Recently, several effective
methods for implementing arbitrary quantum gates have been
reported~\cite{QR_PRL,CSD_PRL,shende_matrix}.
In addition to these constructions, decompositions for certain special
classes of gates have been considered: the local preparation of
quantum states~\cite{shende_vector,stateprep,shende_matrix},
diagonal~\cite{bullock:2004}, and
block-diagonal quantum computations~\cite{hogg}. The important
problem of the implementation of an arbitrary two-qubit gate has also been
recently solved~\cite{shende,whaley_PRL,vatan2,vidal:010301}.
These generic quantum circuit constructions will serve as basic
building blocks for a low-level quantum compiler and
facilitate the optimization of the quantum gate arrays.

The underlying motivation for the pursuit of the optimal quantum
circuit decomposition is decoherence~\cite{decoherence}, which plagues
the practical realizations of quantum computers~\cite{galindo}.
The properties of the quantum compiler and the available gate
primitives strongly influence
the execution time of a quantum algorithm, as is the case with their
classical counterparts. However, owing to the short decoherence times
it is crucial to keep the usage of the computational resources as low as possible,
even for the very first demonstrations of quantum computation.

In this paper, we discuss the properties of uniformly controlled
one-qubit gates which extend the concept of
uniformly controlled rotations introduced in Ref.~\cite{CSD_PRL}.
We give an efficient implementation for these gates in terms of
one-qubit gates and controlled-NOT gates (CNOTs).
Moreover, we observe that our construction can
be implemented effectively also by using only nearest-neighbor gates.
To illustrate the usefulness of the uniformly controlled gates,
we apply them to two examples:
the decomposition of an arbitrary quantum gate and a local
state preparation procedure.
The obtained quantum circuits are quite compact;
in terms of the number of CNOTs involved,
the general gate decomposition is brought on par with the most efficient
currently known general gate decomposition~\cite{shende_matrix}
and somewhat surpasses it in the number of one-qubit gates,
whereas the gate counts required to implement the state preparation
circuit are halved compared to the previous implementations~\cite{stateprep,shende_matrix}.

This paper is organized as follows.
Section~\ref{s:ucu} defines uniformly controlled gates.
In Sec.~\ref{s:qm}, the circuit topology implementing the
uniformly controlled one-qubit gates is constructed.
The implementation is based on the solution of an
eigenvalue equation and is thus cognate to the quantum
multiplexor operation first introduced in Ref.~\cite{shende_matrix}.
In Sec.~\ref{s:examples},
the cosine-sine decomposition (CSD) of an arbitrary
$n$-qubit gate~\cite{CSD_PRL} and a local state preparation
procedure~\cite{stateprep} are improved using this construction.
Finally, in Sec.~\ref{s:nn}, we consider the implementation of the
uniformly controlled one-qubit gates
in a linear chain of qubits with only nearest-neighbor couplings.
Section~\ref{s:jutustelu} is devoted to discussion and a summary of
the results obtained.

\begin{figure}
\includegraphics[width=0.35\textwidth]{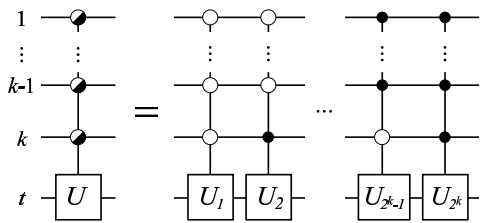}
\caption{\label{fig:ucu}
Uniformly controlled one-qubit gate $\UCU{k}{t}$
stands for a sequence of $k$-fold controlled gates $U_i \in U(2)$,
where $i = 1, \ldots, 2^k$, acting on the
qubit~$t$.
}
\end{figure}

\section{Uniformly controlled gates  \label{s:ucu}}

We define a uniformly controlled one-qubit gate $\UCU{k}{t}$ to be a
sequence of $k$-fold controlled one-qubit gates
in which all the $2^k$ control node configurations are utilized.
All the one-qubit gates in the sequence act on the qubit~$t$, see
Fig.~\ref{fig:ucu}.
We use the symbol $\UCU{k}{t}$ to denote a generic gate of this type,
whereas the full definition of a particular $\UCU{k}{t}$ gate entails
the definition of all the $U(2)$ gates $\{U_i\}_{i=1}^{2^k}$.

Let us now consider the set $G_t(2^n) \subset U(2^{n})$ of
all gates of the form $\UCU{n-1}{t}$.
Each $U \in G_t(2^n)$ is a $2^{n}$-dimensional unitary
operator that can be expressed as a direct sum of
two-dimensional unitary operators~$U_i$, all operating
in subspaces whose basis vectors differ only in the qubit~$t$:
$U~=~\bigoplus_{i=1}^{2^{n-1}}~U_i$.
Since all the operators in $G_t(2^n)$ have identical
invariant subspaces, the set is closed under multiplication and
inversion: assuming that $A, B \in G_t(2^n)$, we have
\begin{align}
\label{eq:kertolaskukaava}
AB     &= \bigoplus_{i=1}^{2^{n-1}} A_i B_i  \in G_t(2^n), \\
A^{-1} &= \bigoplus_{i=1}^{2^{n-1}} A_i^{-1} \in G_t(2^n).
\end{align}
These properties
make $G_t(2^n)$ a subgroup of $U(2^n)$.
We point out that the matrix representations of all the gates
in~$G_t(2^n)$
can be made simultaneously $2\times2$ block-diagonal in the
standard basis using a similarity
transformation, namely a permutation of the qubits, in which the qubit~$t$
is mapped to the qubit~$n$.

As a special case of uniformly controlled one-qubit gates, we define
uniformly controlled rotations~\cite{CSD_PRL}, in which all the two-dimensional
operators $U_i$ belong to the same one-parameter subgroup of $U(2)$,
e.g., the group of rotations around the $z$~axis.
The elements of this particular subgroup are denoted as $\UC{k}{t}{R_z}$.

We extend the notation to accommodate also uniformly controlled multiqubit gates;
by $\UC{k}{\mathcal{T}}{U(2^s)}$ we denote a sequence
of $k$-fold controlled $s$-qubit gates which act
on the set $\mathcal{T}$ of target qubits.

For convenience, we use a shorthand notation for the CNOT and
the below defined two-qubit gate~$D$. The symbol $\CNOT{k}{t}$ is used
to denote a CNOT whose
control and target qubits are the $k$'th and $t$'th, respectively.
Similarly, $D^i_j$ refers to a $D$ gate acting on the qubits~$i$ and~$j$.

\section{Constant quantum multiplexor \label{s:qm}}

\begin{figure}
\includegraphics[width=0.35\textwidth]{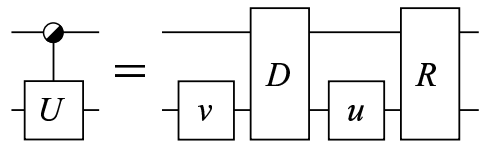}
\caption{\label{fig:qm2} Two-qubit constant quantum multiplexor where
$v$ and $u$ are $SU(2)$ gates, $D$ is a fixed diagonal gate, and
$R$ is an adjustable diagonal gate.}
\end{figure}

Let us start by studying the two-qubit gate
$\UCU{1}{2}$, the matrix representation of which
consists of two unitary $2 \times 2$ blocks.
We show that it can be implemented using the multiplexor circuit
presented in Fig.~\ref{fig:qm2}.
The main difference between the presented construction and the original
quantum multiplexor~\cite{shende_matrix} is that we can effect
the multiplexing operation using a fixed diagonal gate $D$ between the
one-qubit gates. The tradeoff is an additional diagonal gate~$R$
trailing the multiplexor.
The advantage of the proposed construction is that
the fixed gate $D$ can be implemented using a single CNOT,
and in many applications the $R$ gate can be eliminated by
merging it with an adjacent gate.

In matrix form, the implementation of the gate $\UCU{1}{2}$ is
\begin{equation} \label{eq:mplex}
\begin{pmatrix}
a \\
  & b
\end{pmatrix}
=
\underbrace{\begin{pmatrix}
r^\dagger \\
  & r
\end{pmatrix}}_{R}
\, \underbrace{\begin{pmatrix}
u \\
  & u
\end{pmatrix}}_{I \otimes u}
\, \underbrace{\begin{pmatrix}
d \\
  & d^\dagger
\end{pmatrix}}_{D}
\, \underbrace{\begin{pmatrix}
v \\
  & v
\end{pmatrix}}_{I \otimes v},
\end{equation}
where $a$, $b$, $u$ and $v$ are unitary and $r$ and $d$ are diagonal
unitary $2\times2$ matrices.
This yields the matrix equations
\begin{align}
a &= r^\dagger u d v, \\
b &= r u d^\dagger v
\end{align}
or, equivalently,
\begin{align}
\label{eq:ab}
X &:= a b^\dagger = r^\dagger u d^2 u^\dagger r^\dagger, \\
v &= d u^\dagger r^\dagger b = d^\dagger u^\dagger r a. \label{eq:v}
\end{align}

Equation~(\ref{eq:ab}) may be recast into a form reminiscent of an eigenvalue
decomposition:
\begin{equation}
r X r = u d^2 u^\dagger =:  u \Lambda u^\dagger.
\end{equation}
Note that $X$ is fixed by the matrices $a$ and $b$, but $r$ can be
chosen freely.
By diagonalizing the matrix $r X r$, we find the similarity
transformation~$u$ and the eigenvalue matrix $\Lambda = d^2$.
The matrix $v$ is obtained by inserting the results into Eq.~(\ref{eq:v}).

\begin{figure}
\includegraphics[width=0.35\textwidth]{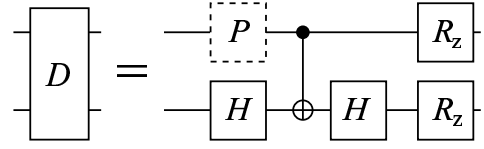}
\caption{\label{fig:d} Elementary gate sequence for the $D$ gate, where
$H$ is the Hadamard gate and $R_z=R_z(\pi/2)$. Gate $P=e^{-i \pi/4}$ is an
adjustment of the global phase and may be omitted.}
\end{figure}

\begin{figure}
\includegraphics[width=0.35\textwidth]{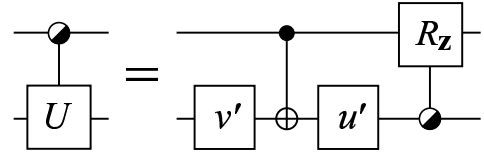}
\caption{\label{fig:qm_rz} Constant quantum multiplexor for two
qubits. Here the $SU(2)$ gates $u'$ and $v'$ include some of the local gates which
transform the CNOT into a $D$ gate. For the implementation of
the gate $\UC{1}{1}{R_z}$, see Fig.~\ref{fig:rzaskel}(a).}
\end{figure}

Since $X \in U(2)$, we may express it using the parametrization
\begin{equation}
X =
\begin{pmatrix}
x_1 & x_2 \\
-\bar{x}_2 & \bar{x}_1
\end{pmatrix}
e^{i \phi / 2},
\end{equation}
where $|x_1|^2 + |x_2|^2 = 1$ and $\det(X) = e^{i \phi}$.
The characteristic polynomial of the matrix $r X r$ is
\begin{equation}
\det(r X r - \lambda I) = \lambda^2 - \lambda \left(r_1^2 x_1 + r_2^2 \bar{x}_1 \right) e^{i \phi /
2} + r_1^2 r_2^2 e^{i \phi}.
\end{equation}
The main result of this section is that for any $X$,
we can find $r$ such that the roots of the polynomial
are two fixed antipodal points on the unit circle in the complex plane.
This is accomplished by choosing $r_i = e^{i \rho_i}$ with
\begin{align}
\rho_1 &= \frac{1}{2} \left( \delta - \frac{\phi}{2} - \arg(x_1) + k \pi \right), \\
\rho_2 &= \frac{1}{2} \left( \delta - \frac{\phi}{2} + \arg(x_1) + m \pi \right).
\end{align}
Above, $k$ and $m$ are arbitrary integers with $k+m$ odd, and $\delta$
is the desired argument for one of the roots $\lambda_i$:
\begin{equation}
\Lambda = d^2 =
\begin{pmatrix}
e^{i \delta} & \\
& -e^{i \delta}
\end{pmatrix}.
\end{equation}
For convenience, let us choose $\delta = \frac{\pi}{2}$. Hence
the diagonal multiplexing gate~$D$ obtains the fixed form
$D~=~e^{i \frac{\pi}{4} \sigma_z \otimes \sigma_z}$.
It can be realized straightforwardly using an Ising-type
Hamiltonian or, alternatively, it can be decomposed into a CNOT
and one-qubit gates as shown in Fig.~\ref{fig:d}.
The resulting diagonal gate~$R$ assumes the form of a uniformly
controlled $z$ rotation in the most significant bit, $\UC{1}{1}{R_z}$.
%
The entire circuit is shown in Fig.~\ref{fig:qm_rz}.

\begin{figure}
\includegraphics[width=0.45\textwidth]{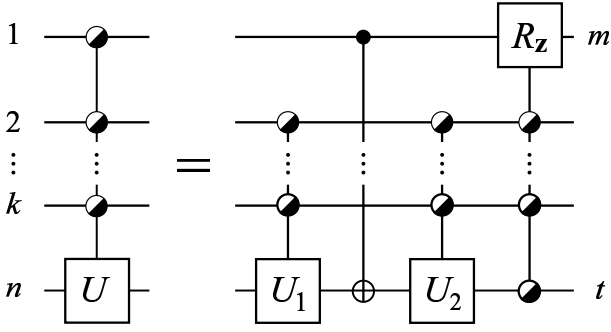}
\caption{\label{fig:askel} Constant multiplexor step for a $k$-fold uniformly
controlled $U(2)$ gate, demultiplexing the qubit~$m$.}
\end{figure}

Now we turn our attention to the decomposition of an arbitrary
$\UCU{k}{t}$ gate, where $k>1$. First we pick one of the control qubits, $m$.
This qubit pairs the two-dimensional invariant subspaces of the gate in a unique fashion.
Hence the method of Eq.~(\ref{eq:mplex})
may be used $2^{k-1}$ times in parallel, which
effectively demultiplexes the chosen control qubit~$m$
of the gate $\UCU{k}{t}$. The operation may be performed using a
single $D^m_t$~gate and a compensating diagonal gate which again assumes the
form of a uniformly controlled $z$~rotation $\UC{k}{m}{R_z}$:
\begin{equation} \label{eq:mplex_step}
\UCU{k}{t} = \UC{k}{m}{R_z} \: \UCU{k-1}{t} \: D^{m}_{t} \: \UCU{k-1}{t}.
\end{equation}
Again, the gate $D^m_t$ may be replaced with a $\CNOT{m}{t}$, see Fig.~\ref{fig:d}, since
the required one-qubit gates may be absorbed into the surrounding gates.
The final form of this step is presented in Fig.~\ref{fig:askel}.

\begin{figure*}
\includegraphics[width=0.95\textwidth]{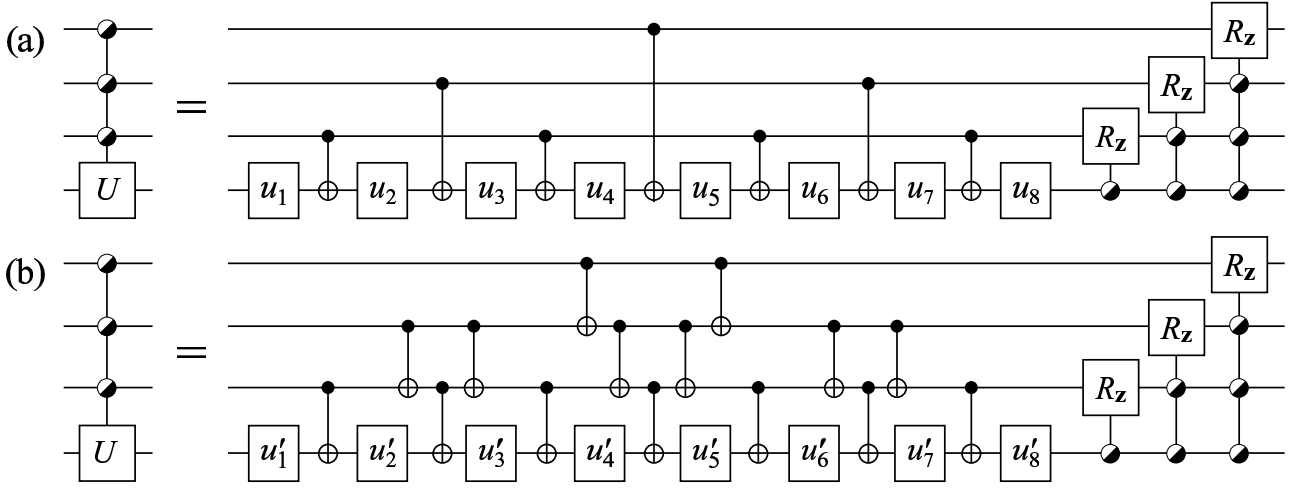}
\caption{\label{fig:3askelta} Implementation of the gate~$\UCU{3}{4}$ using
(a) general CNOTs, (b) only nearest-neighbor CNOTs.
The gates $\{u_i\}$ belong to $SU(2)$.
The alternating sequence of CNOTs and $u_i$ gates is denoted by $\TUCU{3}{4}$.
The rightmost sequence of uniformly controlled $z$ rotations
corresponds to a single diagonal gate, denoted by $\Delta_4$.
For the nearest-neighbor implementation of
uniformly controlled rotations, see Fig.~\ref{fig:nnrz3askelta}.}
\end{figure*}

The decomposition of the gate $\UCU{k}{t}$ can be continued recursively until
only one-qubit gates, CNOTs and uniformly controlled $R_z$~gates are
left.
On the $i$th level of the recursion, there are $2^i$ gates of the type
$\UCU{k-i}{t}$.
The next level of the recursion is obtained by demultiplexing the
control qubit~$j$ in all of these gates.
Given that the leftmost $\UCU{k-i}{t}$ gate is decomposed first,
the resulting $\UC{k-i}{j}{R_z}$ gate, being diagonal, can be commuted
towards the right through the following $D$ gate and merged
with the next $\UCU{k-i}{t}$ gate.
Hence, only the rightmost of the
$\UC{k-i}{j}{R_z}$ gates actually needs to be implemented on each level of the recursion.
The resulting quantum circuit consists of two parts:
an alternating sequence of $2^k$~one-qubit gates and $2^k-1$~CNOTs which we
denote by $\TUCU{k}{t}$, and
a cascade of $k$~distinct uniformly controlled $z$~rotations, which corresponds
to a single diagonal $(k+1)$-qubit gate $\Delta_{k+1}$.
Figure~\ref{fig:3askelta}(a) presents this decomposition for the gate~$\UCU{3}{4}$.

\section{Examples \label{s:examples}}

This section illustrates how the uniformly controlled one-qubit gates
can be applied to efficiently solve two problems:
the decomposition of a general $n$-qubit
gate and the local preparation of an arbitrary quantum state.

\subsection{Cosine-sine decomposition \label{s:csd}}

Recently, we introduced a method~\cite{CSD_PRL} for decomposing a
given general $n$-qubit gate~$U$ into a sequence of
elementary gates using the cosine-sine decomposition~(CSD).
In this approach, the CS decomposition is applied recursively.
Each recursion step decomposes a $k$-fold uniformly controlled
$s$-qubit gate, where $k+s = n$,
into two $(k+1)$-fold uniformly controlled $(s-1)$-qubit
gates and a single $(n-1)$-fold uniformly controlled $y$~rotation:
\begin{align}
\label{eq:csd}
& \UC{k}{\mathcal{T}}{U(2^{s})} = \\
\notag
& \quad \UC{k+1}{
\mathcal{T}\setminus\{m\}
}{U(2^{s-1})} \UC{n-1}{m}{R_y}
\UC{k+1}{
\mathcal{T}\setminus\{m\}
}{U(2^{s-1})}.
\end{align}
Above, $\mathcal{T}$ is the set of $s$ target qubits for the
$U(2^{s})$ gates and $m$ is the operational qubit for the step.
Note that, in this notation, a $U(2^n)$ gate may be denoted as
$\UC{0}{\mathcal{N}}{U(2^{n})}$, where $\mathcal{N}$ is the set of all
the $n$~qubits.
When applied to an arbitrary $n$-qubit gate, the
recursion of Eq.~(\ref{eq:csd}) finally yields the decomposition
\begin{equation}
\label{eq:csdres}
U(2^n) = \UCU{n-1}{n} \prod_{i=1}^{2^{n-1}-1} \UC{n-1}{n-\gamma(i)}{R_y} \UCU{n-1}{n},
\end{equation}
where $\gamma$ is the so-called ruler function, given by Sloane's
sequence A001511~\cite{ruler}.
The order of the noncommuting operators in the product is always taken
to be from left to right.
Note that the $\UC{n-1}{n-\gamma(i)}{R_y}$ gates may as well be considered as
general $\UCU{n-1}{n-\gamma(i)}$ gates.

We continue by decomposing the uniformly controlled gates into
one-qubit gates and CNOTs.
Starting from the last gate in Eq.~(\ref{eq:csdres}),
we write the diagonal part $\Delta_n$ separately:
\begin{equation}
\UCU{n-1}{n} = \Delta_n \TUCU{n-1}{n}.
\end{equation}
The diagonal part $\Delta_n$ can then be merged with the
neighboring $\UC{n-1}{n-1}{R_y}$ gate,
which is transformed into a general gate of type $\UCU{n-1}{n-1}$.
Again, the diagonal part can be separated and
merged into the next gate $\UCU{n-1}{n}$.
Continuing this process sequentially, we finally obtain
\begin{equation}
\label{eq:csdres2}
U(2^n) = \Delta_n \TUCU{n-1}{n} \prod_{i=1}^{2^{n-1}-1}
\TUCU{n-1}{n-\gamma(i)} \TUCU{n-1}{n}.
\end{equation}
This decomposition involves $2^{n} - 1$ gates of type $\TUCU{n-1}{t}$,
each of which takes $2^{n-1}-1$ CNOTs and $2^{n-1}$ one-qubit rotations to
implement.
The final diagonal gate $\Delta_n$ is implemented using the same
construction as in Ref.~\cite{CSD_PRL}.
After eliminating one CNOT and $n$ one-qubit gates,
we obtain a circuit of
$\frac{1}{2}4^n -\frac{1}{2}2^n - 2$ CNOTs and
$ \frac{1}{2}4^n +\frac{1}{2}2^n - n - 1$ one-qubit gates.

Table~\ref{t:comp} presents a comparison between the improved CS
decomposition and the most efficient previously known decomposition,
the NQ decomposition~\cite{shende_matrix}.
The number of CNOTs in the NQ decomposition is from
Ref.~\cite{shende_matrix}.
None of the other results
have been published previously.
\begin{table}[h]
\begin{tabular}{l|c|c}
Gate type & NQ & CS \\
\hline
fixed $U(4)$ & $\frac{1}{2}4^n -\frac{3}{2}2^n + 1$ &
$\frac{1}{2}4^n -\frac{1}{2}2^n - 2$ \\
$R_y, R_z$ & $\frac{9}{8}4^n -\frac{3}{2}2^n + 3$ & $4^n - 1$ \\
or $SU(2)$ & $\frac{17}{24}4^n -\frac{3}{2}2^n - \frac{1}{3}$ &
$\frac{1}{2}4^n +\frac{1}{2}2^n - n - 1$
\end{tabular}
\caption{\label{t:comp} Comparison of the upper bounds for the gate counts required to
implement a general $n$-qubit gate using the NQ
decomposition~\cite{shende_matrix} and the improved CS decomposition.
The fixed $U(4)$ gates may be taken to be CNOTs.
}
\end{table}

\subsection{Local state preparation \label{s:stateprep}}

We have recently addressed~\cite{stateprep} the problem of preparing an arbitrary $n$-qubit
quantum state $\ket{b}_n$ starting from a state $\ket{a}_n$.
The state preparation circuit first transforms the state $\ket{a}_n$ into $\ket{e_1}_n$,
and then, using the same strategy, backwards from $\ket{e_1}_n$ to $\ket{b}_n$.
The $\ket{a}_n$ to $\ket{e_1}_n$ transformation consists of a sequence of gate pairs
\begin{equation}
S_a = \prod_{i=1}^n \left[ \left(\UC{i-1}{i}{R_y}\UC{i-1}{i}{R_z}\right) \otimes
I_{2^{n-i}} \right].
\end{equation}
The effect of the gate pair
$\UC{i-1}{i}{R_y}\UC{i-1}{i}{R_z}$
on the state $\ket{a}_i$ is to nullify half of its elements:
\begin{equation}
\UC{i-1}{i}{R_y}\UC{i-1}{i}{R_z} \ket{a}_i = \ket{a'}_{i-1}\otimes\ket{0}_1.
\end{equation}
Hence, each successive gate pair nullifies half of the
elements of the state vector that have not yet been zeroed, and we have
$S_a \ket{a}_n = \ket{e_1}_n$ up to a global phase.

\begin{figure}
\includegraphics[width=0.35\textwidth]{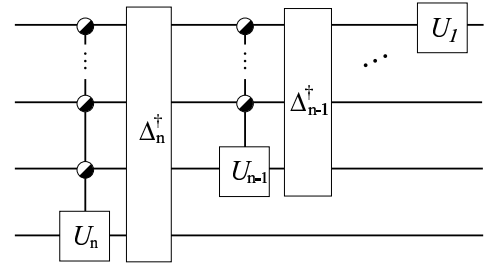}
\caption{\label{fig:vkaanto} Quantum circuit for transforming an arbitrary
$n$-qubit state $\ket{a}_n$ into the standard basis state $\ket{e_1}_n$.
The diagonal gates $\Delta_i^{\dagger}$ exactly cancel the $\Delta_i$ part of the
adjacent $\UCU{i-1}{i}$ gate. The resulting gates are of the form
$\TUCU{i-1}{i}$ which is efficient to implement.}
\end{figure}

Now we note that the pair of gates
$\UC{n-1}{n}{R_y}\UC{n-1}{n}{R_z} = \UCU{n-1}{n}$
may be replaced by the gate
\begin{equation}
\TUCU{n-1}{n}=\Delta_n^{\dagger} \UCU{n-1}{n},
\end{equation}
since the diagonal gate
\begin{equation}
\Delta_n^{\dagger} = \Delta_{n-1}^{0 \, \dagger}\otimes\ket{0}\bra{0}
+\Delta_{n-1}^{1 \, \dagger}\otimes\ket{1}\bra{1}
\end{equation}
does not mix the states;
\begin{align}
\Delta_n^{\dagger} \UCU{n-1}{n} \ket{a}_n &=
\Delta_n^{\dagger} \left(\ket{a'}_{n-1}\otimes\ket{0}_1\right) \notag \\
 &= \left(\Delta_{n-1}^{0 \, \dagger} \ket{a'}_{n-1}\right)\otimes\ket{0}_1 \notag \\
 &= \ket{a''}_{n-1}\otimes\ket{0}_1.
\end{align}
After combining $n-1$ pairs of adjacent
$\UC{k}{k+1}{R_y}\UC{k}{k+1}{R_z}$
gates where $k=1, ..., n-1$
we find that the entire circuit for transforming $\ket{a}$ to $\ket{b}$ requires
$2 \cdot 2^n -2n -2$ CNOTs and $2 \cdot 2^n -n -2$ one-qubit gates.
If $\ket{a}$ or $\ket{b}$ coincides with one of the basis vectors
$\ket{e_i}$, the gate counts are halved in the leading order.
The method presented here
yields a factor-of-two improvement in the gate counts compared to
the previous results~\cite{stateprep,shende_matrix}.
The circuit for this transformation is illustrated in Fig.~\ref{fig:vkaanto}.

\section{Linear chain of qubits with nearest-neighbor couplings \label{s:nn}}

In most of the proposed physical implementations of quantum computers the qubits
are spatially situated
in such a way that only nearest-neighbor interactions are available.
This does not imply that long-range gates are impossible to construct,
but it renders such operations rather hard to implement.
In this section we consider a quantum register consisting of a chain of qubits with only
nearest-neighbor interactions and show that the construction presented
for $\TUCU{k}{t}$ can be translated into an efficient
nearest-neighbor CNOT implementation.
The technique is based on the circuit identity shown in
Fig.~\ref{fig:kaskaadi}.

\subsection{Uniformly controlled one-qubit gates}

\begin{figure}
\includegraphics[width=0.45\textwidth]{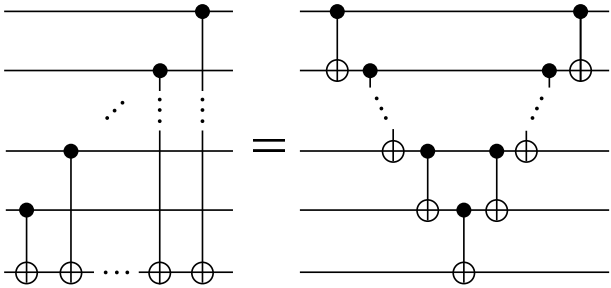}
\caption{\label{fig:kaskaadi} CNOT cascade which can be efficiently
implemented using nearest-neighbor CNOTs.}
\end{figure}

To find the recursion rule for the nearest-neighbor implementation of
a uniformly controlled one-qubit gate,
we modify the recursion rule expressed in Eq.~(\ref{eq:mplex_step})
by inserting an identity in the form of a CNOT cascade and its
inverse, a similar cascade,
into the circuit next to the multiplexing gate $\CNOT{m}{t}$.
The cascades consist of the gates $\CNOT{i}{t}$, where $i$ runs over the qubits
connecting the qubits $m$ and~$t$.
One of the cascades is absorbed into the following
$\UCU{j}{t}$. The remaining cascade, together with the multiplexing
CNOT, can be efficiently implemented using nearest-neighbor CNOTs as
illustrated in Fig.~\ref{fig:kaskaadi}.

The complexity of the nearest-neighbor implementation depends on the
relative order of the target and control qubits, and the order in
which the control qubits are demultiplexed.
Since the number of nearest-neighbor CNOTs required increases linearly with the
distance between the control and target qubits of the multiplexing CNOT,
we first demultiplex the control qubits that are furthest apart
from the target.
Let us assume that a $\TUCU{n-1}{t}$ gate acts on a chain of $n$
consequent qubits.
If $n \ge 5$, it is advantageous to use a sequence of swap
gates to move the target qubit next to the center of the chain before the
operation and back after it. A swap gate can be realized using three
consecutive CNOTs.
Taking this into account, a $\TUCU{n-1}{t}$ gate can be implemented using at most
\begin{equation} \label{eq:ucu-nncnots}
C_{U(2)}(n, s) = \frac{5}{6} 2^{n} +2n-6s
- \left\{\begin{matrix}
\frac{1}{3}, \quad \text{$n$ even} \\
\frac{5}{3}, \quad \text{$n$ odd}
\end{matrix} \right.
\end{equation}
nearest-neighbor CNOTs, where $s = 1, \ldots, \lceil\frac{n}{2} \rceil$
is the distance of the target qubit~$t$ from the end of the chain.
Figure~\ref{fig:3askelta}(b) depicts the resulting circuit for the
case $n=4$ and $s=1$.

\begin{figure}
\includegraphics[width=0.45\textwidth]{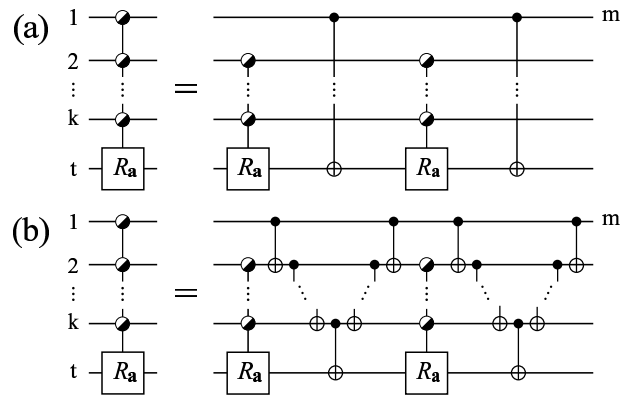}
\caption{\label{fig:rzaskel} Recursion step for decomposing a
uniformly controlled rotation using
(a) CNOTs
(b) nearest-neighbor CNOTs, applied to the qubit~$m$.
Note that the circuit diagrams may also be mirrored horizontally.
}
\end{figure}

\begin{figure*}
\includegraphics[width=0.95\textwidth]{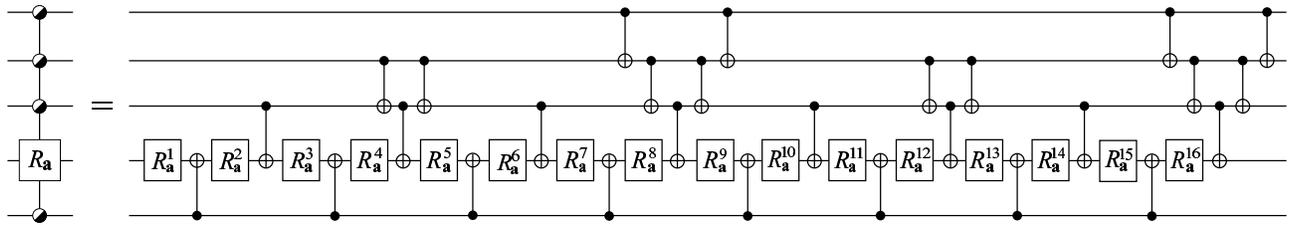}
\caption{\label{fig:nnrz3askelta} Implementation of a uniformly
controlled ${\bf a}$ rotation using nearest-neighbor CNOTs.}
\end{figure*}

Now consider a $k$-fold uniformly controlled
rotation gate $\UC{k}{t}{R_{\bf a}}$, where the rotation axis ${\bf a}$
is perpendicular to the $x$~axis. It can be decomposed using the
recursion step presented in Fig.~\ref{fig:rzaskel}(b).
To minimize the CNOT count, we mirror at each recursion step
the circuit of the latter uniformly controlled
gate, which results in
the cancellation of two nearest-neighbor CNOT cascades.
For the same reason as in the previous paragraph,
the recursion step is first applied to
the control qubits furthest apart from the target.
The implementation for the gate $\UC{n-1}{t}{R_{\bf a}}$ requires at most
\begin{equation} \label{eq:ucr-nncnots}
C_{R}(n, s) = \frac{5}{6} 2^{n} +3n-6s 
-\left\{\begin{matrix}
\frac{4}{3}, \quad \text{$n$ even} \\
\frac{5}{3}, \quad \text{$n$ odd}
\end{matrix} \right.
\end{equation}
nearest-neighbor CNOTs.
Figure~\ref{fig:nnrz3askelta} displays an example circuit for the case
$n=5$ and $s=2$.

\subsection{Cosine-sine decomposition}

The decomposition of an arbitrary $n$-qubit gate
is achieved exactly as in Sec.~\ref{s:csd}, but now
the order in which the CSD steps of Eq.~(\ref{eq:csd}) are applied to the qubits affects
the final gate count. As seen in Eq.~(\ref{eq:ucu-nncnots}), it is
favorable to have the target qubit of a uniformly controlled
one-qubit gate as close to the center of the chain
as possible. Consequently, we start the decomposition from
the ends of the qubit chain, moving alternatingly towards the center.
In this fashion, a general $n$-qubit gate can be implemented using at most
\begin{equation}
C_{U}(n) = \frac{5}{6} 4^n - n 2^n -2n
+\left\{\begin{matrix}
\frac{5}{6} 2^n -\frac{5}{3}, \quad \text{$n$ even} \\
\frac{1}{2} 2^n -\frac{1}{3}, \quad \text{$n$ odd}
\end{matrix} \right.
\end{equation}
nearest-neighbor CNOTs.

\subsection{Local state preparation}

With the help of the results derived above, the
implementation of the general state preparation circuit using
nearest-neighbor gates is straightforward.
We follow the reasoning of
Sec.~\ref{s:stateprep} and simply replace
the $\TUCU{i-1}{i}$ gates
with their nearest-neighbor counterparts,
using the decomposition derived in the beginning of this section.
We find that the implementation of the state preparation circuit
requires at most
\begin{equation}
C_{\text{SP}}(n) = \frac{10}{3} 2^n + 2 n^2 - 12n 
+\left\{\begin{matrix}
\frac{14}{3}, \quad \text{$n$ even}  \\
\frac{10}{3}, \quad \text{$n$ odd} 
\end{matrix} \right.
\end{equation}
nearest-neighbor CNOTs.

\section{\label{s:jutustelu} Discussion}

In this paper we have studied the properties and the utilization of
uniformly controlled one-qubit gates.
We have derived a recursive circuit topology which implements
an arbitrary $k$-fold uniformly controlled one-qubit gate using at most
$2^{k}$~one-qubit gates, $2^{k}-1$~CNOTs and a single diagonal $(k+1)$-qubit gate.
This construction is especially efficient if the gate is to implemented
only up to a diagonal, e.g. when the phase factors of each basis vector
can be freely chosen. We have also shown that this kind of freedom
appears in the implementation of an arbitrary $n$-qubit quantum gate
and in the rotation of an arbitrary state vector into another.
The leading-order complexity of the circuit for an arbitrary
$n$-qubit gate is
$\frac{1}{2}4^n$ CNOTs and an equal number of one-qubit gates, which
are the lowest gate counts reported.

The techniques presented above are also amenable to experimental
realizations of a quantum computer in which the quantum register
consists of a one-dimensional chain of qubits with nearest-neighbor
interactions.
For example, the number of the nearest-neighbor
CNOTs in the presented decomposition of an $n$-qubit gate is in the
leading order $\frac{5}{6} 4^n$,
which is appreciably below the lowest previously reported value of
$\frac{9}{2} 4^n$~\cite{shende_matrix}. Furthermore, the structure of
the nearest-neighbor circuit allows several gate operations to be executed
in parallel, which may further reduce the execution time of the algorithm.

In Ref.~\cite{CSD_PRL}, it was speculated that the gate count of the
quantum CSD could be reduced by combining adjacent uniformly
controlled rotations into single uniformly controlled one-qubit gates,
which was realized in this paper. To further reduce the number of
CNOTs in the circuit, also the control nodes of the CNOTs should be used to
separate the one-qubit gates carrying the degrees of freedom. However,
uniformly controlled one-qubit gates cannot be used as the sole basic
building blocks of the circuit in this kind of a construction.


\begin{acknowledgments}
This research is supported by the Academy of Finland (project
No. 206457, ``Quantum Computing'').
VB and MM thank the Finnish Cultural Foundation,
JJV and MM the Jenny and Antti Wihuri Foundation,
and JJV the Nokia Foundation for financial support.
\end{acknowledgments}

\bibliography{qc}

\end{document}